\documentclass[aps,showpacs,preprintnumbers,amsmath, amssymb]{revtex4}

\usepackage{float}
\usepackage{graphics,epsfig}
\usepackage{graphicx}
\usepackage{dcolumn}
\usepackage{bm}

\begin{document}
\baselineskip=0.8 cm
\title{{\bf  Dark energy model with higher derivative of Hubble parameter}}
\author{Songbai Chen}
\email{csb3752@163.com} \affiliation{Institute of Physics and
Department of Physics,
Hunan Normal University,  Changsha, Hunan 410081, P. R. China \\
Key Laboratory of Low Dimensional Quantum Structures and Quantum
Control (Hunan Normal University), Ministry of Education, P. R.
China.}

\author{Jiliang Jing }
\email{jljing@hunnu.edu.cn}
 \affiliation{Institute of Physics and
Department of Physics,
Hunan Normal University,  Changsha, Hunan 410081, P. R. China \\
Key Laboratory of Low Dimensional Quantum Structures and Quantum
Control (Hunan Normal University), Ministry of Education, P. R.
China.}

\vspace*{0.2cm}
\begin{abstract}
\baselineskip=0.6 cm
\begin{center}
{\bf Abstract}
\end{center}

In this letter we consider a dark energy model in which the energy
density is a function of the Hubble parameter $H$ and its derivative
with respect to time $\rho_{de}=3\alpha
\ddot{H}H^{-1}+3\beta\dot{H}+3\gamma H^2$. The behavior of the dark
energy and the expansion history of the Universe depend heavily on
the parameters of the model $\alpha$, $\beta$ and $\gamma$. It is
very interesting that the age problem of the well-known three old
objects can be alleviated in this models.

\end{abstract}

\maketitle
\newpage
\vspace*{0.2cm}

It has been strongly confirmed that our Universe is undergoing an
accelerated expansion by many observations including the Type Ia
supernova (SN) \cite{A1}, cosmic microwave background (CMB)
\cite{A2} and large scale structure (LSS) \cite{A3}, and so on. The
late time cosmic acceleration challenges our understanding of the
standard models of gravity and particle physics. Within the
framework of Einstein's general relativity, the Universe is supposed
to be filled with dark energy to explain this observed phenomena.
The dark energy is an exotic energy component with negative pressure
and the experiments have indicated that today it constitutes about
$70\%$ of present total cosmic energy. However, so far, the nature
of dark energy is still unclear.

It seems that the preferred candidate for dark energy is the
cosmological constant \cite{1a}, which is a term that can be added
to Einstein's equations. This term acts like a perfect fluid with an
equation of state $\omega=-1$, and the energy density is associated
with quantum vacuum. Recent investigations indicate that the
cosmological constant is consistent with observational data.
However, it is very difficult to understand in the modern field
theory since the vacuum energy density is far below the value
predicted by any sensible quantum field theory. Moreover, it has
also been plagued by the so-called coincidence problem, namely,
``why are the vacuum and matter energy densities of precisely the
same order today?". In order to eliminate these problems, a lot of
the dynamical scalar fields, such as quintessence \cite{2a},
k-essence \cite{3a}, phantom \cite{4a} and quintom field
\cite{5a,5a1}, have been put forth as an alternative of dark energy.
The another possible way of explaining the cosmic accelerated
expansion is that Einstein's theory should be modified, for example
the $f(R)$ theory \cite{6a} and the DGP model \cite{7a}.

It is well known that the holographic principle \cite{qx1} plays an
important role in the black hole and string theory, which is based
on the fact in quantum gravity, the entropy of a system scales not
with its volume, but with its surface area $L^2$. Inspired by the
holographic principle,  A. Cohen \textit{et al} \cite{Aco} suggested
that the vacuum energy density is proportional to the Hubble scale
$l_H \sim H^{-1}$. In this model, both the fine-tuning and
coincidence problems can be alleviated, but it can not explain the
cosmic accelerated expansion because that the effective equation of
state for such vacuum energy is zero. Recently, M. Li \cite{Li}
proposed that the future event horizon of the Universe to be used as
the characteristic length $l$. This holographic dark energy model
not only presents a reasonable value for dark energy density, but
also leads to an acceleration solution for the cosmic expansion. In
fact, the choice of the characteristic length $l$ is not the unique
for the holographic dark energy model. Gao \textit{et al} \cite{Gao}
assumed that the length $l$ is giving by the the inverse of Ricci
scalar curvature, i.e., $|R|^{-1/2}$, which is the so-called
holographic Ricci dark energy model. It is argued that this model
can solve the coincidence problem entirely. Thus, the properties of
such holographic Ricci dark energy have been investigated widely
recently \cite{Cai,hr1,hr2,hr3,hr4,hr5,hr6,hr7,hr8,hr81}. L. N.
Granda \textit{et al} \cite{hr9,hr10} proposed  a modified Ricci
dark energy model in which the density of dark energy is a function
of the Hubble parameter $H$ and its derivative with respect to time
$\dot{H}$. However, all of these models have been plagued with the
age problem of the well-known three old objects, LBDS $53W091$
$(z=1.55, t=3.5 Gyr)$ \cite{ag1}, LBDS 53W069 ($z=1.43,t=4.0 Gyr)$
\cite{ag2} and APM 08279+5255 $(z=3.91, t=2.1 Gyr)$ \cite{ag3}. In
this letter, we present a generalized dark energy model in which
density of dark energy contains the second order derivative with
respect to time $\ddot{H}$ and find that the age problem of the old
objects above can be alleviated.

It is well known that in the Einstein general relativity theory, a
flat universe is described by the standard Friedmann equation
\begin{eqnarray}
H^2=\frac{1}{3}\rho.\label{fd1}
\end{eqnarray}
This means that the total density $\rho$ of the Universe is a
function of the Hubble parameter $H$. Since the dark energy occupies
almost $70\%$ of the content of the universe today, it is rational
to assume that the density of dark energy is a function of the
Hubble parameter $H$ and its derivatives with respect to time. For
the mathematical simplicity, we here assume that the density of dark
energy contains the Hubble parameter $H$, the first order and the
second order derivatives (i.e, $\ddot{H}$ and $\dot{H}$). The
concrete expression of the density of dark energy is given by
\begin{eqnarray}
\rho_{de}=3\alpha \ddot{H}H^{-1}+3\beta \dot{H}+3\gamma
H^2,\label{de}
\end{eqnarray}
where $\alpha$, $\beta$ and $\gamma$ are three arbitrary
dimensionless parameters. The main reason we choice such a form
(\ref{de}) for the density of dark energy is that the Friedman
equation (\ref{fd1}) can be rewritten as a second order differential
equation with constant coefficients (see Eq.(\ref{weif}) below) and
we can obtain an analytical general solution for it, which is very
convenient for us to study the properties of the model in the
following calculations. Moreover, we introduce the inverse of Hubble
parameter (i.e, $H^{-1}$) in the first term in the density of dark
energy so that the dimensions of the terms in the equation
(\ref{de}) are identical. When $\alpha=0$, it can be reduce to the
modified Ricci dark energy model \cite{hr9,hr10}.  Since possessing
an extra free parameter $\alpha$, the model (\ref{de}) is more
general than the modified Ricci dark energy. The similar dark energy
models have also been studied in \cite{hf1,hf2}.

Setting the variable $x=\ln{a}$ and substituting the density of dark
energy (\ref{de}) into Eq.(\ref{fd1}), the Friedman equation can be
written as
\begin{eqnarray}
\frac{\alpha}{2}\frac{d^2h^2}{dx^2}+\frac{\beta}{2}\frac{dh^2}{dx}+(\gamma-1)h^2+\Omega_{m0}e^{-3x}=0,\label{weif}
\end{eqnarray}
where $h=H/H_0$, $\Omega_{m0}=\rho_{m0}/(3H^2_0)$ and $H_0$ is the
Hubble constant. Neglecting the contribution from the radiation, the
general solution of the differential equation (\ref{weif}) can
expressed by
\begin{eqnarray}
h^2=\Omega_{m0}e^{-3x}+f_0e^{-\frac{\beta-\sqrt{\beta^2-8\alpha(\gamma-1)}}{2\alpha}x}+f_1e^{-\frac{\beta+\sqrt{\beta^2-8\alpha(\gamma-1)}}{2\alpha}x}
+\frac{9\alpha-3\beta+2\gamma}{2-9\alpha+3\beta-2\gamma}\Omega_{m0}e^{-3x},\label{h}
\end{eqnarray}
where $f_0$ and $f_1$ are integration constants. From the initial
condition we find that $f_0$ and $f_1$ satisfy
\begin{eqnarray}
f_0+f_1+\frac{2}{2-9\alpha+3\beta-2\gamma}\Omega_{m0}=1.\label{in1}
\end{eqnarray}
From this equation we obtain that the integration constants $f_0$
and $f_1$ can not be decided exactly. In other word, there exists a
free integration constant between $f_0$ and $f_1$. Here we only
consider the cases $f_1=0$, $f_0=0$ and $f_1=f_0$ for simplicity,
and then we study the properties of the dark energy model (\ref{de})
and check the age problem of three old objects.

Let us first consider the case $f_1=0$. It is very easy to obtain
that the density and pressure of dark energy can be expressed by
\begin{eqnarray}
\rho_{de}=H^2_0\bigg[f_0e^{-\frac{\beta-\sqrt{\beta^2-8\alpha(\gamma-1)}}{2\alpha}x}+\frac{9\alpha-3\beta+2\gamma}{2-9\alpha+3\beta-2\gamma}\Omega_{m0}e^{-3x}\bigg],
\end{eqnarray}
and
\begin{eqnarray}
p_{de}=-\rho_{de}-\frac{1}{3}\frac{d\rho_{de}}{dx}=-\frac{6\alpha-\beta+\sqrt{\beta^2-8\alpha(\gamma-1)}}{6\alpha}f_0H^2_0
e^{-\frac{\beta-\sqrt{\beta^2-8\alpha(\gamma-1)}}{2\alpha}x},
\end{eqnarray}
respectively. Here
\begin{eqnarray}
f_0=\frac{2-9\alpha+3\beta-2\gamma-2\Omega_{m0}}{2-9\alpha+3\beta-2\gamma}.
\end{eqnarray}
Thus, the equation of state is
\begin{eqnarray}
\omega=\frac{p_{de}}{\rho_{de}}=-\frac{(6\alpha-\beta+\sqrt{\beta^2-8\alpha(\gamma-1)})(2-9\alpha+3\beta-2\gamma-2\Omega_{m0})
e^{\frac{6\alpha-\beta+\sqrt{\beta^2-8\alpha(\gamma-1)}}{2\alpha}x}}{6\alpha[(9\alpha-3\beta+2\gamma)\Omega_{m0}+
(2-9\alpha+3\beta-2\gamma-2\Omega_{m0})
e^{\frac{6\alpha-\beta+\sqrt{\beta^2-8\alpha(\gamma-1)}}{2\alpha}x}]}.\label{w1}
\end{eqnarray}
\begin{figure}[ht]
\begin{center}
\includegraphics[width=8cm]{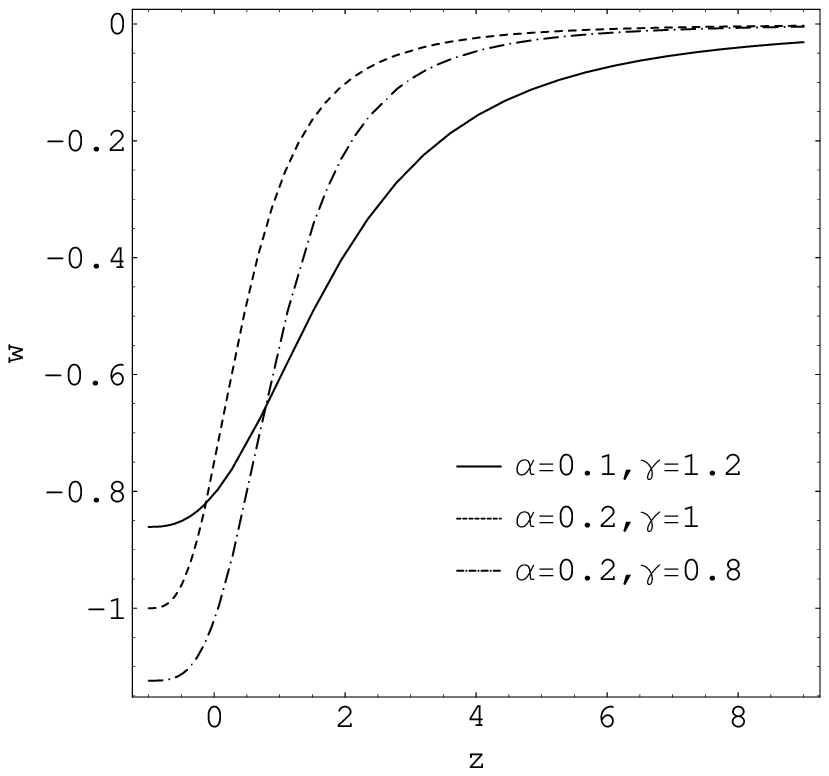}
\caption{The change of the equation of state $\omega$ with the
redshift $z$ for different $\alpha$ and $\gamma$.  Here we set
$\beta=1$ and $\Omega_{m0}=0.27$. }
\end{center}
\end{figure}
Obviously, the equation of state of dark energy depends on the
parameters $\alpha$, $\beta$, $\gamma$, $\Omega_{m0}$ and the
variable $x$. From Eq.(\ref{w1}), it is easy to find that when
$9\alpha-3\beta+2\gamma=0$ the equation of state of dark energy is a
constant
$\omega=-1+\frac{\beta-\sqrt{\beta^2-8\alpha(\gamma-1)}}{6\alpha}$.
Especially, when $\gamma=1$ and, the model (\ref{de}) can recover
the cosmological constant \cite{1a}. As $\alpha(\gamma-1)>0$
($\alpha(\gamma-1)<0$), it is corresponded to the constant $\omega$
quintessence \cite{2a} (phantom \cite{4a}) model. For the case
$9\alpha-3\beta+2\gamma\neq0$, it describes the model in which the
equation of state of dark energy is variable with the time. The
behavior of dark energy (\ref{de}) depends on the values of
$\alpha$, $\beta$, $\gamma$ and $\Omega_{m0}$. From the Fig. (1), we
find that for chosen $\alpha$, $\beta$ and $\gamma$ the dark energy
(\ref{de}) has the behaviors of ``quintom-like" field (such as
$\alpha=0.2$, $\beta=1$, $\gamma=0.8$) \cite{5a,5a1} and of
``quintessence-like" matter ( such as $\alpha=0.1$, $\beta=1$,
$\gamma=1.2$) \cite{2a}, which are similar to that of the modified
holographic Ricci dark energy \cite{Gao,hr9,hr10}.
\begin{figure}[ht]
\begin{center}
\includegraphics[width=8cm]{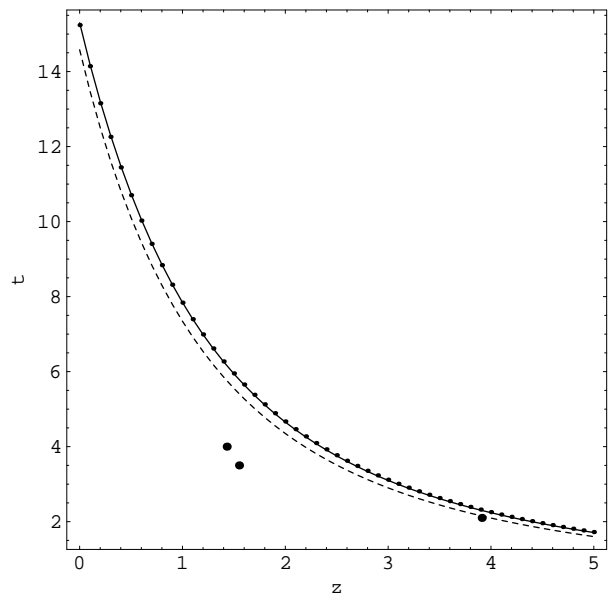}
\caption{The change of the age of Universe with the redshift $z$.
The solid, dotted and dashed lines denote the cases ($\alpha=0.2$,
$\beta=2.4$, $\gamma=0.8$), ($\alpha=0.1$, $\beta=2.2$,
$\gamma=0.9$) and ($\alpha=0.3$, $\beta=2.6$, $\gamma=1$),
respectively. The large points denote the ages of the three old
objects LBDS $53W091$ $(z=1.55, t=3.5 Gyr)$, LBDS 53W069
($z=1.43,t=4.0 Gyr)$ and APM 08279+5255 $(z=3.91, t=2.1 Gyr)$. Here
we set $\Omega_{m0}=0.27$ and $h_0=0.78$.}
\end{center}
\end{figure}

The evolution of age of the Universe is described as
\begin{eqnarray}
t=\frac{1}{H_0}\int^{-\ln{(1+z)}}_{-\infty}\frac{dx}{h},
\end{eqnarray}
where $x=\ln{a}$ and $h=H/H_0$. From the observations of the Hubble
Space Telescope Key project, the present Hubble parameter is
constrained to be $H^{-1}_{0}=9.776h^{-1}_0$ , where $0.64 < h_0 <
0.80$. In Fig.(2) we plotted the curves of age of the Universe for
different values of $\alpha$, $\omega$ and $\Omega_{m0}$. We also
check the ages problem of several old objects, LBDS $53W091$
$(z=1.55, t=3.5 Gyr)$ \cite{ag1}, LBDS 53W069 ($z=1.43,t=4.0 Gyr)$
\cite{ag2} and APM 08279+5255 $(z=3.91, t=2.1 Gyr)$ \cite{ag3} and
find that for chosen parameter the ages problem can be alleviated in
this case.

Now let us consider the second case $f_0=0$. Similarly, the density
and pressure of dark energy can be written as
\begin{eqnarray}
\rho_{de}=H^2_0\bigg[f_1e^{-\frac{\beta+\sqrt{\beta^2-8\alpha(\gamma-1)}}{2\alpha}x}+\frac{9\alpha-3\beta+2\gamma}{2-9\alpha+3\beta-2\gamma}\Omega_{m0}e^{-3x}\bigg],
\end{eqnarray}
and
\begin{eqnarray}
p_{de}=-\rho_{de}-\frac{1}{3}\frac{d\rho_{de}}{dx}=-\frac{6\alpha-\beta-\sqrt{\beta^2-8\alpha(\gamma-1)}}{6\alpha}f_1H^2_0
e^{-\frac{\beta+\sqrt{\beta^2-8\alpha(\gamma-1)}}{2\alpha}x},
\end{eqnarray}
respectively. The integration constants $f_1$
\begin{eqnarray}
f_1=\frac{2-9\alpha+3\beta-2\gamma-2\Omega_{m0}}{2-9\alpha+3\beta-2\gamma}.
\end{eqnarray}
\begin{figure}[ht]
\begin{center}
\includegraphics[width=8cm]{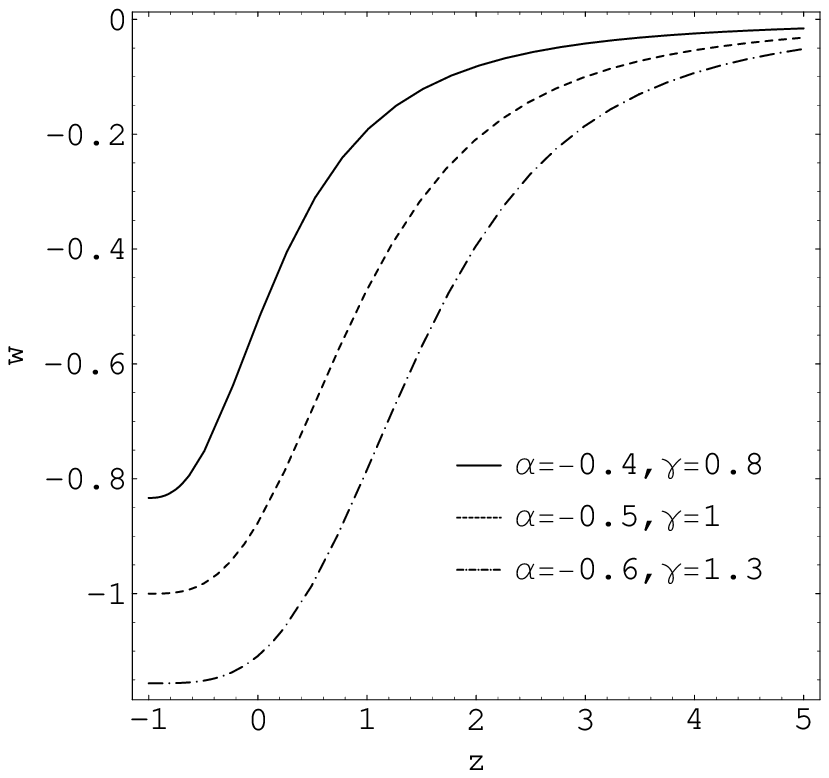}
\caption{The change of the equation of state $\omega$ with the
redshift $z$ for different $\alpha$ and $\gamma$.  Here we set
$\beta=-1$ and $\Omega_{m0}=0.27$. }
\end{center}
\end{figure}
Thus, the equation of state is
\begin{eqnarray}
\omega=\frac{p_{de}}{\rho_{de}}=-\frac{(6\alpha-\beta-\sqrt{\beta^2-8\alpha(\gamma-1)})(2-9\alpha+3\beta-2\gamma-2\Omega_{m0})
e^{3x}}{6\alpha[(9\alpha-3\beta+2\gamma)\Omega_{m0}e^{\frac{\beta+\sqrt{\beta^2-8\alpha(\gamma-1)}}{2\alpha}x}+
(2-9\alpha+3\beta-2\gamma-2\Omega_{m0})e^{3x}]}.\label{w2}
\end{eqnarray}
Similarly, as $9\alpha-3\beta+2\gamma=0$ the equation of state of
dark energy is a constant
$\omega=-1+\frac{\beta+\sqrt{\beta^2-8\alpha(\gamma-1)}}{6\alpha}$.
When $\alpha<-2/9$ and $\beta=(9\alpha+2)/3$, we find that the
equation of state $\omega=-1$. It is shown in Fig. (3) that  the
dark energy (\ref{de}) has the behaviors of ``quintom-like" field
and of ``quintessence-like" matter for different values of $\alpha$,
$\beta$ and $\gamma$ in this case. Moreover, as in the case I, the
age problem of old high redshift objects can be alleviated in the
model (\ref{de}).
\begin{figure}[ht]
\begin{center}
\includegraphics[width=8cm]{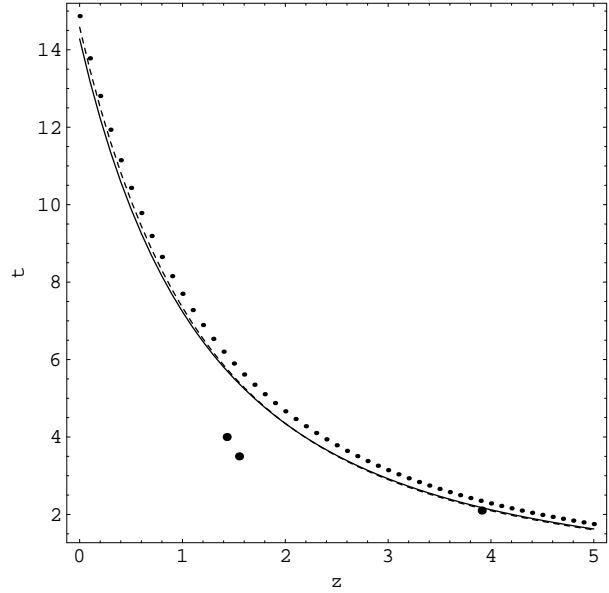}
\caption{The change of the age of Universe with the redshift $z$.
The solid, dotted and dashed lines denote the cases ($\alpha=-0.9$,
$\beta=-1$, $\gamma=0.9$), ($\alpha=-1$, $\beta=-1$, $\gamma=0.92$)
and ($\alpha=-0.9$, $\beta=-1$, $\gamma=1$), respectively. The large
points denote the ages of the three old objects LBDS $53W091$
$(z=1.55, t=3.5 Gyr)$, LBDS 53W069 ($z=1.43,t=4.0 Gyr)$ and APM
08279+5255 $(z=3.91, t=2.1 Gyr)$. Here we set $\Omega_{m0}=0.27$ and
$h_0=0.78$.}
\end{center}
\end{figure}

Now we consider the case III: $f_0=f_1$. From the initial condition
(\ref{in1}), we obtain that
\begin{eqnarray}
f_0=f_1=\frac{2-9\alpha+3\beta-2\gamma-2\Omega_{m0}}{2(2-9\alpha+3\beta-2\gamma)}.
\end{eqnarray}
Repeating the previous operation, we find that the density, pressure
and the equation of state of dark energy are
\begin{eqnarray}
\rho_{de}=H^2_0\bigg[f_1\bigg(e^{-\frac{\beta+\sqrt{\beta^2-8\alpha(\gamma-1)}}{2\alpha}x}+
e^{-\frac{\beta-\sqrt{\beta^2-8\alpha(\gamma-1)}}{2\alpha}x}\bigg)
+\frac{9\alpha-3\beta+2\gamma}{2-9\alpha+3\beta-2\gamma}\Omega_{m0}e^{-3x}\bigg],
\end{eqnarray}
\begin{eqnarray}
p_{de}=-\rho_{de}-\frac{1}{3}\frac{d\rho_{de}}{dx}
&=&\frac{f_1H^2_0}{6\alpha}\bigg[\bigg(\beta+\sqrt{\beta^2-8\alpha(\gamma-1)}-6\alpha\bigg)e^{-\frac{\beta+\sqrt{\beta^2-8\alpha(\gamma-1)}}{2\alpha}x}
\nonumber\\
&&+\bigg(\beta-\sqrt{\beta^2-8\alpha(\gamma-1)}-6\alpha\bigg)e^{-\frac{\beta-\sqrt{\beta^2-8\alpha(\gamma-1)}}{2\alpha}x}\bigg],
\end{eqnarray}
and
\begin{eqnarray}
\omega=\frac{p_{de}}{\rho_{de}}=\frac{f_1\bigg[\bigg(\beta+\sqrt{\beta^2-8\alpha(\gamma-1)}-6\alpha\bigg)e^{-\frac{\beta+\sqrt{\beta^2-8\alpha(\gamma-1)}}{2\alpha}x}
+\bigg(\beta-\sqrt{\beta^2-8\alpha(\gamma-1)}-6\alpha\bigg)e^{-\frac{\beta-\sqrt{\beta^2-8\alpha(\gamma-1)}}{2\alpha}x}
\bigg]}{6\alpha\bigg[f_1\bigg(e^{-\frac{\beta+\sqrt{\beta^2-8\alpha(\gamma-1)}}{2\alpha}x}+
e^{-\frac{\beta-\sqrt{\beta^2-8\alpha(\gamma-1)}}{2\alpha}x}\bigg)
+\frac{9\alpha-3\beta+2\gamma}{2-9\alpha+3\beta-2\gamma}\Omega_{m0}e^{-3x}\bigg]},\nonumber\\
\label{w3}
\end{eqnarray}
\begin{figure}[ht]
\begin{center}
\includegraphics[width=8cm]{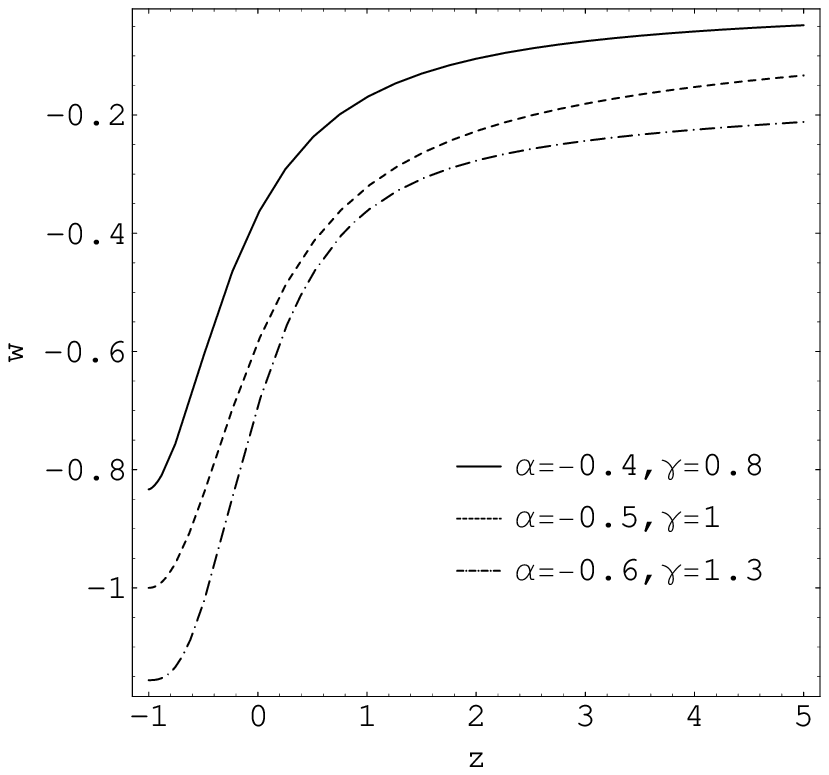}
\caption{The change of the equation of state $\omega$ with the
redshift $z$ for different $\alpha$ and $\gamma$.  Here we set
$\beta=-1$ and $\Omega_{m0}=0.27$. }
\end{center}
\end{figure}
\begin{figure}[ht]
\begin{center}
\includegraphics[width=8cm]{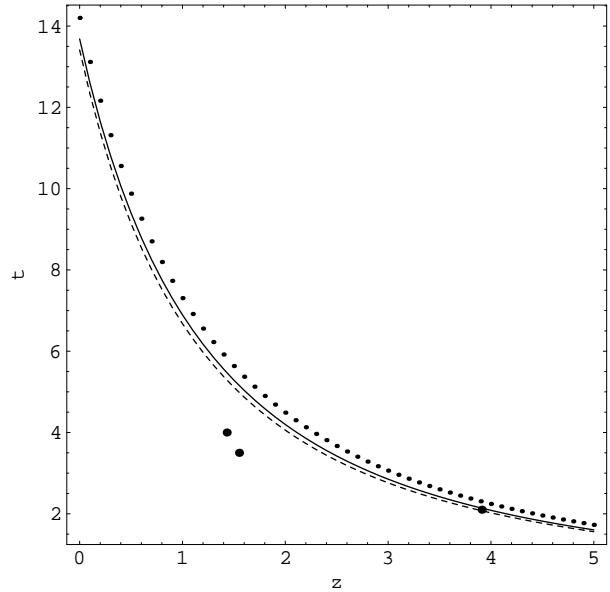}
\caption{The change of the age of Universe with the redshift $z$.
The solid, dotted and dashed lines denote the cases ($\alpha=-0.9$,
$\beta=-1$, $\gamma=0.9$), ($\alpha=-1$, $\beta=-1$, $\gamma=0.92$)
and ($\alpha=-0.9$, $\beta=-1$, $\gamma=1$), respectively. The large
points denote the ages of the three old objects LBDS $53W091$
$(z=1.55, t=3.5 Gyr)$, LBDS 53W069 ($z=1.43,t=4.0 Gyr)$ and APM
08279+5255 $(z=3.91, t=2.1 Gyr)$. Here we set $\Omega_{m0}=0.27$ and
$h_0=0.78$.}
\end{center}
\end{figure}
respectively. Similarly, as $2-9\alpha+3\beta-2\gamma=0$ and
$\beta^2-8\alpha(\gamma-1)=0$, the equation of state of dark energy
becomes a constant $\omega=-1+\frac{\beta}{6\alpha}$. It is easy to
obtain that as $\gamma=1,\beta=0$ it reduces to cosmological
constant. In the general case, the equation of state of dark energy
is variable with the time which is shown in Fig.(5). We also examine
the age problem of old high redshift objects and find it can be
alleviated in this case.

In summary, we studied a dark energy model with higher derivative of
Hubble parameter. This model can be reduced to the dark energy with
the constant $\omega$, Ricci-like dark energy models, and so on. The
behavior of the dark energy and the expansion history of the
Universe depend heavily on the parameters of the model $\alpha$,
$\beta$ and $\gamma$. We check the age problem of the three old
objects for three special cases in the model and find it can be
alleviated for the chosen parameters. It implies that this kind of
studies can help us to understand more about dark energy.

\acknowledgments
This work was partially supported by the National
Natural Science Foundation of China under Grant No.10875041; the
Scientific Research Fund of Hunan Provincial Education Department
Grant No.07B043 and the construct program of key disciplines in
Hunan Province. J. L. Jing's work was partially supported by the
National Natural Science Foundation of China under Grant No.10675045
and No.10875040; and the Hunan Provincial Natural Science Foundation
of China under Grant No.08JJ3010.


\begin{thebibliography}{}
\baselineskip=0.6 cm

\bibitem{A1}A. G. Riess \textit{et al}, Astron. J. {\bf 116},  1009 (1998) [arXiv:astro-ph/9805201];

P. de Bernardis \textit{et al}, Nature {\bf 404},  955 (2000);

S. Perlmutter \textit{et al}, Astrophys. J. {\bf 517},  565 (1999)
[arXiv:astro-ph/9812133]; Astrophys. J. {\bf 598},  102 (2003).

\bibitem{A2} D. N. Spergel \textit{et al}, Astrophys. J.
Suppl. {\bf170},  377 (2007)[arXiv:astro-ph/0603449];

E. Komatsu \textit{et al},  Astrophys. J. Suppl. {\bf180}, 330
(2009) [arXiv:0803.0547 [astro-ph]].

\bibitem{A3} M. Tegmark \textit{et al}, Phys. Rev. D {\bf69}, 103501 (2004)
[arXiv:astro-ph/0310723];

J. K. Adelman-McCarthy \textit{et al}, Astrophys. J. Suppl.
{\bf175}, 297 (2008) [arXiv:0707.3413 [astro-ph]].


\bibitem{1a} S. Weinberg, Rev. Mod. Phys. {\bf 61},  1 (1989);

V. Sahni and A. Starobinsky, Int. J. Mod. Phy. D {\bf 9}, 373
(2000);

P. J. E. Peebles and B. Ratra, Rev. Mod. Phys. {\bf 75},  559
(2003);

T. Padmanabhan, Phys. Rept.  {\bf 380}, 235 (2003);

E. J. Copel and M. Sami and S. Tsujikawa, Int. J. Mod. Phys.  D {\bf
15}, 1753 (2006).


\bibitem{2a} B. Ratra and P. J. E. Peebles, Phys. Rev. D {\bf37},  3406 (1988);

P. J. E. Peebles  and B. Ratra,  Astrophys. J. {\bf325},  L17
(1988);

C. Wetterich,  Nucl. Phys. B {\bf302}, 668 (1988);

R. R. Caldwell, R. Dave and P. J. Steinhardt, Phys. Rev. Lett.
{\bf80}, 1582 (1998) ;

I. Zlatev, L. Wang and P. J. Steinhardt, Phys. Rev. Lett. {\bf82},
896 (1999);

M. Doran and J. Jaeckel,  Phys. Rev. D. {\bf66}, 043519 (2002).

\bibitem{3a} C. A.  Picon, T. Damour and V. Mukhanov, Phys. Lett. B
{\bf458}, 209 (1999);

T. Chiba, T. Okabe and M. Yamaguchi, Phys. Rev. D {\bf62},  023511
(2000).

\bibitem{4a} R. R. Caldwell,  Phys. Lett. B {\bf545},  23 (2002);

B. McInnes, J. High Energy Phys. {\bf0208}, 029 (2002);

S. Nojiri and S. D. Odintsov, Phys. Lett. B {\bf562},  147 (2003);

L. P. Chimento and R. Lazkoz, Phys. Rev. Lett. {\bf91},  211301
(2003);

B. Boisseau, G. Esposito-Farese, D. Polarski, Alexei A.
Starobinsky,Phys. Rev. Lett. {\bf85},  2236 (2000);

R. Gannouji, D. Polarski, A. Ranquet, A. A. Starobinsky, JCAP
{\bf0609}, 016 (2006), [astro-ph/0606287].

\bibitem{5a} B. Feng, X. L. Wang and X. M. Zhang, Phys. Lett. B {\bf607}, 35 (2005)
[arXiv:astro-ph/0404224];

Z. K. Guo, Y. S. Piao, X. M. Zhang and Y. Z. Zhang, Phys. Lett. B
{\bf608},  177 (2005) [arXiv:astro-ph/0410654];

\bibitem{5a1} Y. F. Cai, M. Z. Li, J. X. Lu, Y. S. Piao, T. T. Qiu and X. M. Zhang, Phys. Lett. B {\bf651}, 1 (2007)
[arXiv:hep-th/0701016];

Y. F. Cai, H. Li, Y. S. Piao and X. M. Zhang, Phys. Lett. B
{\bf646}, 141 (2007) [arXiv:gr-qc/0609039];

W. Zhao and Y. Zhang, Phys. Rev. D {\bf73}, 123509 (2006)
[arXiv:astro-ph/0604460];

S. G. Shi, Y. S. Piao and C. F. Qiao, arXiv: 0812.4022 [astro-ph];

H. Mohseni Sadjadi and M. Alimohammadi, Phys. Rev. D {\bf74}, 043506
(2006) [arXiv:gr-qc/0605143];

M. R. Setare and E. N. Saridakis, arXiv:0807.3807 [hep-th];

M. R. Setare and E. N. Saridakis, JCAP {\bf0809}, 026 (2008)
[arXiv:0809.0114 [hep-th]];

E. Elizalde, S. Nojiri and S. D. Odintsov, Phys. Rev. D {\bf 70},
043539 (2004) [arXiv: hep-th/0405034].

\bibitem{6a} S. Capozziello, Int. J. Mod. Phys. D {\bf11}, 483 (2002);

T. P. Sotiriou, V. Faraoni, arXiv:0805.1726;

A. A. Starobinsky, Phys. Lett. B {\bf91}, 99 (1980);

S. Nojiri and S. D. Odintsov, arXiv: 0807.0685 [hep-th];

S. Nojiri and S. D. Odintsov, Int. J. Geom. Meth. Mod. Phys. {\bf
4}, 115 (2007) arXiv: hep-th/0601213.

\bibitem{7a} G. Dvali, G. Gabadadze and M. Porrati, Phys. Lett. B {\bf485}, 208 (2000).

\bibitem{qx1} G'.t Hooft, arXiv:gr-qc/9310026;

L. Susskind, J. Math. Phys. {\bf36}, 6377 (1995)
[arXiv:hep-th/9409089].

\bibitem{Aco} A. Cohen, D. Kaplan and A. Nelson, Phys. Rev. Lett.
{\bf82}, 4971 (1999);

P. Horava and D. Minic, hep-th/hepth/ 0001145, Phys. Rev. Lett. {\bf
85}, 1610 (2000);

S. Thomas, Phys. Rev. Lett. {\bf 89},  081301 (2002).

\bibitem{Li} M. Li, Phys. Lett. B {\bf603},  1 (2004).

\bibitem{Gao} C. Gao, F. Q. Wu, X. Chen and Y. G. Shen, Phys. Rev. D
{\bf79}, 043511 (2009) [arXiv:0712.1394].

\bibitem{Cai} R. G. Cai, B. Hu and Y. Zhang, arXiv: 0812.4504 [hep-th].

\bibitem{hr1} Q. G. Huang and M. Li, JCAP {\bf0408}, 013 (2004)
[arXiv:astro-ph/0404229]; JCAP {\bf0503}, 001 (2005)
[arXiv:hep-th/0410095];

\bibitem{hr2}X. Zhang and F. Q. Wu, Phys. Rev. D {\bf72}, 043524 (2005)
[arXiv:astro-ph/0506310]; Phys. Rev. D {\bf76}, 023502 (2007)
[arXiv:astro-ph/0701405];

\bibitem{hr3} Q. G. Huang and Y. G. Gong, JCAP {\bf0408}, 006 (2004)
[arXiv:astro-ph/0403590];

Z. Chang, F. Q. Wu and X. Zhang, Phys. Lett. B {\bf633}, 14 (2006)
[arXiv:astro-ph/0509531];

M. R. Setare, J. Zhang and X. Zhang, JCAP {\bf0703}, 007 (2007)
[arXiv:gr-qc/0611084];

Y. G. Gong, Phys. Rev. D {\bf70}, 064029 (2004)
[arXiv:hep-th/0404030];

B. Wang, E. Abdalla and R. K. Su, Phys. Lett. B {\bf611}, 21 (2005)
[arXiv:hep-th/0404057];

J. Zhang, X. Zhang and H. Liu, Phys. Lett. B {\bf651}, 84 (2007)
[arXiv:0706.1185 [astroph]];

Y. Z. Ma and X. Zhang, Phys. Lett. B {\bf661}, 239 (2008)
[arXiv:0709.1517 [astro-ph]];

X. Wu, R. G. Cai and Z. H. Zhu, Phys. Rev. D {\bf77}, 043502 (2008)
[arXiv:0712.3604 [astro-ph]];

B. Chen, M. Li and Y. Wang, Nucl. Phys. B {\bf774}, 256 (2007)
[arXiv:astro-ph/0611623]

\bibitem{hr4} B. Wang, Y. G. Gong and E. Abdalla, Phys. Lett. B {\bf624}, 141 (2005)
[arXiv:hep-th/0506069];

B. Wang, C. Y. Lin and E. Abdalla, Phys. Lett. B {\bf637}, 357
(2006) [arXiv:hep-th/0509107];

B. Wang, C. Y. Lin, D. Pavon and E. Abdalla, Phys. Lett. B {\bf662},
1 (2008) [arXiv:0711.2214 [hep-th]]

\bibitem{hr5}C. J. Feng, arXiv:0806.0673 [hep-th];  Phys. Lett. B {\bf670},
231 (2008) [arXiv:0809.2502 [hep-th]]; arXiv:0810.2594 [hep-th];

L. Xu, W. Li and J. Lu, arXiv:0810.4730 [astro-ph];

K. Y. Kim, H. W. Lee and Y. S. Myung, arXiv: 0812.4098 [gr-qc].

\bibitem{hr6} X. Zhang, Phys. Lett. B {\bf648}, 1 (2007) [arXiv:astro-ph/0604484];

X. Zhang, Phys. Rev. D {\bf74}, 103505 (2006)
[arXiv:astro-ph/0609699];

\bibitem{hr7}X. Zhang, arXiv: 0901.2262; arXiv: 0904.0045.

\bibitem{hr8} S. Nojiri and S. D. Odintsov, GRG {\bf 38}, 1285 (2006)
hep-th 0506212.

\bibitem{hr81} C. J. Feng and X. Z. Li, arXiv: 0905.0527.

\bibitem{hr9} L. N. Granda and A. Oliveros, Phys. Lett. B {\bf 669}, 275
(2008); Phys. Lett. B {\bf 671}, 199 (2009);  arXiv: 0901.0561.

\bibitem{hr10} L. N. Granda, arXiv: 0811.4103; arXiv: 0812.1596.

\bibitem{ag1} J. Dunlop \textit{et al}, Nature {\bf 381}, 581 (1996).

\bibitem{ag2} H. Spinrad \textit{et al}, Astrophys. J {\bf 484}, 581 (1999).

\bibitem{ag3} D. Jain and A. Dev, Phys. Lett. B {\bf 633}, 436 (2006)
[astro-ph/0509212].


\bibitem{hf1} S. Nojiri and S. D. Odintsov, Phys. Rev. D {\bf 72}, 023003 (2005) hep-th
0505215;

S. Capozziello, V. Cardone, E. Elizalde, S. Nojiri and S. D.
Odintsov,  Phys. Rev. D {\bf 73}, 043512 (2006)  astro-ph: 0508350.

\bibitem{hf2} K. Bamba, S. Nojiri and S. D. Odintsov, JCAP {\bf 0810}, 045 (2008)
arXiv:0807.2575 [hep-th].

\end{thebibliography}
\end{document}